\documentstyle[11pt, epsf, a4, cite]{article}

\newcommand{\mquad}{\!\!\!\!\!\!}
\newcommand{\mqquad}{\!\!\!\!\!\!\!\!\!\!\!\!}

\newcommand{\gl}{\;\;=\;\;}
\newcommand{\tgl}{\!=\!}
\newcommand{\tGl}[2]{\,$#1 \!=\! #2 $\,}

\newcommand{\pl}{\;+\;}
\newcommand{\mi}{\;-\;}
\newcommand{\fr}[2]{\frac{#1}{#2}}
\newcommand{\tfr}[2]{\mbox{\Large $\frac{#1}{#2}$}}
\newcommand{\nfr}[2]{\frac{\mbox{$#1$}}{\mbox{$#2$}}}

\newcommand{\be}{\begin{equation}}
\newcommand{\ee}{\end{equation}}
\newcommand{\bea}{\begin{eqnarray}}
\newcommand{\eea}{\end{eqnarray}}
\newcommand{\bib}[1]{\vspace{-1ex}\bibitem{#1}}

\newcommand{\bc}{\begin{center}}
\newcommand{\ec}{\end{center}}

\newcommand{\bt}[1]{\begin{tabular}{#1}}
\newcommand{\et}{\end{tabular}}

\renewcommand{\(}{\left(}
\renewcommand{\)}{\right)}
\renewcommand{\[}{\left[}
\renewcommand{\]}{\right]}

\newcommand{\eq}[1]{(eq.~(\ref{#1}))}
\newcommand{\eqn}[1]{eq.~(\ref{#1})}

\newcommand{\bm}[1]{\mbox{\mathversion{bold} $\! #1$}}
\newcommand{\bmind}[1]{\mbox{\scriptsize \mathversion{bold} $\! #1$}}

\newcommand{\eins}{\mbox{$1\!\!\!\;{\rm l}$}}
\newcommand{\ph}{\varphi}
\newcommand{\PH}{\mbox{\mathversion{bold} $\!\Phi$}}

\newcommand{\del}{\partial}

\newcommand{\Tr}{{\rm Tr}\,}
\newcommand{\cvec}[2]{{#1 \choose #2}}
\newcommand{\calO}{{\cal O}}

\newcommand{\calM}{{\cal M}}
\newcommand{\ra}{\rightarrow}

\newcommand{\mdrsq}{\overline{m}_3^2}

\newcommand{\ladr}{\bar\lambda_3}

\def\gev{{\rm GeV}}

\def\lsim{\mbox{\raisebox{0.05cm}[0cm][0cm]{$<$}\hspace{-0.42cm}
          \raisebox{-0.12cm}[0cm][0cm]{$\sim$}}}
\def\gsim{\mbox{\raisebox{0.05cm}[0cm][0cm]{$>$}\hspace{-0.42cm}
          \raisebox{-0.12cm}[0cm][0cm]{$\sim$}}}

\let\3=\ss

\begin{document}

\begin{titlepage}
\begin{flushright}
HD--THEP--96--53\\
DO--TH~96/26\\
hep-ph/9612450\\
December 23, 1996
\end{flushright}
\vspace{1.5cm}
\begin{center}
{\LARGE\bf  Non-perturbative correlation masses \\[1ex]
                 in the hot electroweak phase}\\
\vspace{1cm}
{\bf
Hans G\"unter Dosch $^1$\\
\vspace{.1cm}
Jochen Kripfganz $^2$\\
\vspace{.1cm}
Andreas Laser $^3$\\
\vspace{.1cm}
Michael G.~Schmidt $^1$\\
}
\vspace{1cm}
$^1$ Institut  f\"ur Theoretische Physik,
Universit\"at Heidelberg\\
Philosophenweg 16,
D-69120 Heidelberg, FRG\\[2ex]
$^2$ SoftAS GmbH\\
Sch\"onauer Str.~14, D-04430 Lindennaundorf, FRG\\[2ex]
$^3$ Institut f\"ur Physik,
Theorie III\\
Universit\"at Dortmund,
D-44221 Dortmund, FRG\\
\vspace{1.5cm}
{\bf Abstract}\\
\end{center}
The effective action describing the long range fluctuations in the 
high temperature phase of the electroweak 
standard theory is a strongly coupled SU(2)-Higgs-model in three 
dimensions.  We outline in detail a model in which the spatial correlation  
scales in this phase are calculated as inverse relativistic bound state masses. 
Selection rules for these states are derived. The correlation masses
are calculated by evaluating the bound state Green's function. 
The scalar-scalar-potential and its influence on the masses
is investigated. 
The predictions for the correlation masses agree very well with the lattice data 
available now.
\vfill
{\small
{\parindent0cm \hrulefill\\
e-mail addresses:}\\[0.5ex]
\begin{tabular}{l} 
  H.G.Dosch@thphys.uni-heidelberg.de\\
  jk@softas-gmbh.l.eunet.de\\
  laser@hal1.physik.uni-dortmund.de\\
  M.G.Schmidt@thphys.uni-heidelberg.de
\end{tabular}
}
\end{titlepage}

\section{Introduction}

The perturbative treatment of the electroweak standard model (SM) leads to a 
spec\-ta\-cu\-lar\-ly good agreement with experiments. It works because the 
Higgs mechanism renders the non-Abelian gauge bosons massive.
This remains true in the Higgs phase (broken phase) of the 
SM at high temperatures for small Higgs masses,  
as has been discussed carefully in the last years 
\cite{FarakosEA,ArnoldEs,FodorHe,Wir2,Laine1,KajantieEA1}.
But due to the rising Higgs-plasma-mass at very high temperatures 
$T>T_c\sim 100\gev$ a new hot (unbroken) phase of the electroweak theory 
was argued to exist \cite{KirzhnitsLi} in which according to perturbative 
calculations the gauge bosons are massless. A first order phase transition
between the two phases could have important effects, in particular the 
possibility of a baryogenesis \cite{KuzEA} at this state has been discussed
extensively \cite{RuScha}.
It now is practically excluded for the 
experimentally allowed range of $m_H\gsim 60\gev$ but variants of the standard  
model in particular the supersymmetric standard model are under discussion.

If one wants to discuss the phase transition 
in detail -- critical bubbles, sphaleron transitions, etc.\ -- one needs
the effective action (or at least the effective potential) of the theory.
However, for vanishing or small Higgs vacuum expectation values the infrared
problem comes up like in QCD and we expect non-perturbative effects.
In the case of large $m_H\gsim 70\gev$, where the phase transition 
starts to fade away, this non-perturbative regime might cover even 
the broken phase at the temperatures of the phase transition \cite{BeWe}. 

At high temperatures the non-zero Matsubara-modes are heavy 
and the IR-sensitive part of the theory is  an effective 
three dimensional theory of the zero modes whose parameters 
are obtained from the original theory by matching a set of 
amplitudes in 4 and 3 dimensions \cite{FarakosEA,KajantieEA1}.
The three dimensional action can be simplified further by integrating
out the massive zero modes of the time component of the gauge field.
In the high temperature expansion higher derivative terms are suppressed 
and neglecting electroweak mixing 
one ends up with the three dimensional SU(2)-Higgs model. 
An appropriate way to treat such an IR-sensitive theory where non-perturbative
effects are expected is Monte Carlo simulations on the lattice like in 
QCD. This is simpler than in the latter case because the fermions  
have already been integrated out as non-zero Matsubara-modes.
Such calculations have been performed by several groups 
\cite{IlgenfEA1,GuertlerEA1,GuertlerEA2,PhilipsenEA,KarschEA,KajantieEA2}.
They confirmed that there are confinement effects to be discussed
in detail later on. An effective 3-dimensional gauge coupling rising
in the infrared was also argued for in the framework of exact 
renormalization group equations \cite{Wett}.

In this paper we will discuss the spatial correlation scales 
of the physical states, i.e.\ of the corresponding gauge 
invariant operators, in the 
3-dimensional SU(2)-Higgs-model. The simple-minded perturbative picture 
would predict long range correlations because of the massless gauge bosons,
but the neglect of IR-effects is obviously wrong. Based on 1-loop
gap equations it was argued that one has just another
Higgs-phase \cite{BuPh}. The predicted vector-boson-mass 
is much smaller than the 
one calculated in the gauge invariant lattice calculations (but curiously
not much different from (Landau) gauge fixed lattice results \cite{KarschEA}).
In a recent letter \cite{Wir3} we proposed a model for two-dimensional 
bound states of light constituents in 
the hot electroweak phase. The Green's functions of these 
bound states have
been evaluated following a method which was developed by 
Simonov \cite{Simonov3} in 4-dimensional QCD.
In this paper we elaborate the model in detail. The evaluation of the 
Green's function is generalized and reorganized. The bound state correlation  
masses are first calculated analytically for a linear scalar-scalar-potential.
Several modifications of the potential and their influences on the
correlation masses are discussed. 
The parameters of the potential are then fixed from 
lattice calculations of the Wegner-Wilson-loop.
Finally we can compare our predictions for higher correlations
with recent gauge invariant lattice calculations 
\cite{PhilipsenEA,GuertlerEA2}.

Chapter 2 introduces the effective 3-dimensional SU(2)-Higgs-model.
Chapter 3 discusses the possible bound states. The corresponding 
Green's functions are derived in chapter 4. They are evaluated in the
next chapter. In chapter 6 we calculate the correlation masses for 
a linear potential, discuss modifications of this potential and 
their influence on the masses. The parameters of our model 
are fixed in chapter 7. Chapter 8 gives the comparison of
our results with lattice data. The intercept of the potential is
investigated in chapter 9.
Finally we present our conclusions.

\section{The SU(2)-Higgs-model in three dimensions}

At high temperatures the SM can effectively be described by
the three dimensional SU(2)-Higgs-model \cite{KajantieEA1} with
the Lagrangian
\be
{\cal L} \gl
\fr14 \, F_{ij}^a F_{ij}^a \pl
(D_i\phi)^\dagger (D_i\phi) \pl
m_3^2\,\phi^{\dagger}\phi\pl
\lambda_3(\phi^{\dagger}\phi)^2
\quad,\qquad\quad
D_i=\del_i \;-\; g_3 A_i^a\fr{\tau^a}{2} 
\quad.
\label{V3}
\ee
Even variants of the SM, e.g\ the minimal supersymmetric standard model, 
can be described by 
this effective theory in a large part of the parameter space \cite{Mikko4}. 
The squared mass $m_3^2$, the quartic coupling $\lambda_3$ 
and  the three dimensional gauge coupling $g_3$ depend on the 
parameters of the fundamental theory and on the temperature.
The quartic coupling has the mass-dimension 1, while $g_3$ has the 
mass-dimension 1/2.

The non-perturbative aspects we are interested in are dominated by the 
gauge boson sector. In contrast to a four dimensional theory this 
sector has a natural mass scale given by the gauge coupling. It is
therefore natural to express dimensioned quantities in units 
of powers of $g_3$. The parameters of the model can then be represented 
by the two dimensionless quotients
\be\label{contpar}
\ladr\gl\fr{\lambda_3}{g_3^2}
\qquad\qquad\qquad{\rm and} \qquad\qquad\qquad
\mdrsq \gl\fr{m_3^2}{g_3^4}
\quad,
\ee
sometimes called $x$ and $y$ in the literature \cite{KajantieEA1}.
The mass $m_3$ depends in general on the renormalization scale.  
We work, however, with fixed values of $\mdrsq$ and hence at a 
fixed scale. This is similar to lattice calculations where $\mdrsq$
is fixed at the scale given by the lattice constant. 

In order to uncover the full symmetry of the SU(2)-Higgs-model
we replace the Higgs-doublet $\phi(x)$
by a $2\times 2$-matrix-field $\Phi(x)$ via 
\be
\phi\gl \cvec{\phi^+}{\phi^0}
 \gl \bm{\Phi}\, \cvec{0}{1}
\qquad\quad{\rm with}\qquad\quad 
\bm{\Phi} \gl \Phi_0 \,\eins_{2\!\times\! 2} \pl i\, \Phi_i \,\tau^i 
\quad,
\label{skalarMatrix}
\ee
where the $\tau^a$ are the Pauli matrices.
Due to the identity 
\,$ \Tr\!\bm{\Phi}^\dagger\!\bm{\Phi}\tgl 2\,\phi^\dagger\phi$\,
the potential \eq{V3} depends only on this trace. 

Expressing the gauge field by the usual 
matrix notation 
\,$\bm{A}_i \tgl A_i^a \;\tfr{\tau^a}{2}$\,
the action can be written as
\be
S_3 \gl \int d^3x \;\left\{\;\fr12\,
\Tr\!\bm{F}_{ij} \bm{F}_{ij}\pl
\fr12\,\Tr(\bm{D}_i\bm{\Phi})^{\dagger}(\bm{D}_i\bm{\Phi})\pl
V_3\(\fr12\Tr\!\bm{\Phi}^\dagger\bm{\Phi}\)\;\right\}\quad.
\label{S3matrix}
\ee
It is invariant under the transformation
\bea
\bm{\Phi}(x) &\rightarrow & \bm{U}\!(x)\; \bm{\Phi}(x)\; \bm{V} \quad,\\[0.5ex]
\bm{D}_i(x) &\rightarrow& 
         \bm{U}\!(x)\; \bm{D}_i(x) \;\bm{U}\!(x)^\dagger\quad,\\[0.5ex]
\bm{F}_{ij}(x) &\rightarrow& 
         \bm{U}\!(x)\; \bm{F}_{ij}(x) \;\bm{U}\!(x)^\dagger\quad.  
\eea
$\bm{U}\!(x)$ is an element of the 
local gauge group ${\rm SU(2)_G}$ and depends in 
general on $x$, while $V$ is a $x$-independent element of the 
isospin group ${\rm SU(2)_I}$.
Hence the first index $\alpha$ of the  
matrix $\bm{\Phi}_{\alpha a}$ refers to the gauge group and the second index $a$
to the global $SU(2)_I$.

has a first gauge index $\alpha$ and
a second isospin index $a$.

\section{Bound states of two scalars}\label{scalarboundstates}

In the 3-dimensional SU(2)-Higgs-model like in all gauge theories 
in principle everything has to be expressed in terms of gauge invariant
quantities. For the well known formulation of the Higgs mechanism 
for generating massive gauge bosons this just implies a rather trivial 
reformulation in terms of gauge invariant objects (unitary gauge).
In the  case of a confining theory such a gauge invariant description 
is standard and absolutely mandatory. Indeed some time ago the 
4-dimensional SU(2)-Higgs model was a prominent model for describing
an effective electroweak  interaction of composite $W$, Higgs, quarks
and leptons.
But it is not clear just from a gauge invariant formulation whether
one is in the Higgs or in the confinement phase. Contrary to the 
case of the original Abbot-Farhi-model for compositeness in electroweak 
interaction \cite{AbFa,SchDoKr,MatveevEA} we argue here that in the 
3-dimensional case of the 
hot electroweak theory one really has a non-perturbative theory with 
genuine composite states in 2+1 dimensions. 
In this we differ 
from reference \cite{BuPh} where a Higgs-mechanism also in this phase
is advocated.

\paragraph*{Nonlocal operators}

The nonlocal operators corresponding to the bound state of two fundamental
scalars $\PH$ are either
\be\label{nlsing}
\Tr\, \PH(x)^\dagger \bm{T}(x,\bar{x}) \PH(\bar{x}) 
\ee
or
\be\label{nltrip}
\quad \Tr\, \PH(x)^\dagger \bm{T}(x,\bar{x}) \PH(\bar{x})\, \tau^i  \quad.
\ee
The matrix $\bm{T}(x,\bar{x})$ with gauge-indices 
is the link-operator (also called gauge field transporter) defined by
\be\label{linkop}
\bm{T}(x,\bar{x}) \gl {\cal P} \exp\!\( \,- \,i g_3 \int^{\bar{x}}_x 
\bm{A}_k\, dz_k\,\) \quad,
\ee
where ${\cal P}$ denotes the path-ordering operator along some 
given path between $x$ and $\bar x$.
The operator of \eqn{nlsing} is an isospin-singlet, 
the operator of \eqn{nltrip}
is an isospin-triplet; both are gauge singlets. 
Since they are non-local all possible angular momentum states contribute.

The local interpolating fields
\be\label{interpolfield}
\Tr\, \PH(x)^\dagger \PH(x) 
\qquad\qquad{\rm and}\qquad\qquad
\Tr\, \PH(x)^\dagger \bm{D}_k \PH(x)\, \tau^i  
\ee
are included in these operators
in the limit \,$\bar{x}\!\ra\!x$.
The index of the vector field corresponds to the spatial
direction of the link operator. 
Note that $\Tr\, \PH(x)^\dagger \PH(x)\, \tau^i$   
vanishes while $\Tr\, \PH(x)^\dagger \bm{D}_k \PH(x)$
is a total derivaitve.

\paragraph*{Selection rules}

The local limit shows that the isospin-singlet operator overlaps with 
scalar states while the isospin-triplet operator overlaps with vector
states. The obvious question arises if there are scalar isospin-triplets
or vector isospin-singlets as well. 
Let us first assume that the path between $x$ and $\bar x$ is a straight line.
In coordinate gauge the operator $\bm{T}(x,\bar{x})$ is then  
equal to the $\eins$-matrix. 
The gauge-invariant operators 
(\ref{nlsing}) and (\ref{nltrip}) can be evaluated in this gauge. 
One gets
\be
\Tr\, \PH(-r)^\dagger \PH(r)\, \calO_I  
\qquad\qquad  {\rm with} \qquad\qquad  
\calO_I \in \{ \eins_{2\times 2} , \tau^j \} 
\quad.
\ee
Here we have changed the coordinate system; the origin is now 
at the center of the bound state. 

The projection on parity-even and parity-odd states respectively is
\be
\fr12 \; \Tr\, \( \PH(-r)^\dagger \PH(r) \;\pm\; 
                  \PH(r)^\dagger \PH(-r)\, \) \calO_I 
\quad.
\ee
Expressing $\PH(r)$ by real fields $\Phi_0$ and $\Phi_i$ as 
in \eqn{skalarMatrix} the parity-even states become 
\be
\Big(\, \Phi_0(-r) \Phi_0(r) \pl \Phi_i(-r) \Phi_i(r) \,\Big) 
\; \Tr\;\eins_{2\times 2} \;\calO_I
\ee
while the parity-odd states are
\be
i \, \Big(\, \Phi_0(-r) \Phi_i(r) \mi \Phi_i(-r) \Phi_0(r) 
\mi \Phi_k(-r) \Phi_l(r) \epsilon_{kli} \, \Big) 
 \; \Tr\;\tau^i \;\calO_I 
\quad.
\ee
One sees the parity-even isospin-triplets (with $\calO_I\!=\!\tau^j$) and
parity-odd isospin-singlets (with $\calO_I\!=\!\eins_{2\times 2}$) vanish. 
Hence in particular no scalar isospin-triplets and no vector-isospin-singlets
are allowed.

If the path between $x$ and $\bar x$ is not a straight line there may be
an overlap with states which do no respect this selection rule.
Nevertheless, there is no local interpolating field composed of two scalar 
fields alone
which corresponds to
these states in  the limit \,$\bar x\!\ra\! x$.
The interpolating fields in that case would contain at least an additional
gauge boson field $F_{ij}$ and thus correspond to hybrid states in QCD.
The latter ones are several hundred MeV (order of magnitude of the 
``gluon constituent'' mass) heavier than the corresponding pure 
quark-antiquark states. We assume that a similar mechanism holds also in 
our case. 

The derivation of the selection rules given above can be generalized to
``blocked'' operators which are used in lattice investigations.

\section{Green's functions and correlation masses}

In order to get some insight into the dynamical structure of the
electroweak theory above the critical temperature and to compare with 
QCD in the confining phase we have proposed in reference \cite{Wir3}
a bound state model with a potential whose parameters were determined 
by comparing with results of lattice simulations 
for Wegner-Wilson-loops. Since the fundamental scalars might be very 
light one has to treat the problem with relativistic kinematics.
For that we adopted the 
method of Simonov \cite{Simonov3} to 
our case. 
We reorganize and generalize our 
calculations of reference \cite{Wir3} in this paper. 
In the described model the correlation masses
are determined from the exponential falloff of the Green's function.

\paragraph*{The Green's function of a fundamental scalar}

We start from the Green's function $\bm{G}(x,y,\!\bm{A})_{a \alpha b \beta}$ 
that transforms the matrix field $\PH(x)_{a\alpha}$ into $\PH(y)_{b\beta}$.
Neglecting the scalar self-interaction it has in the 
worldline formalism \cite{SchmSchu} the form
\bea\label{gfktfs1}
\mqquad&&
\bm{G}(x,y,\!\bm{A})_{a \alpha b \beta}\\[1ex]
\mqquad&&
=\;\;
\int_0^\infty \!\!ds \exp\!\(-\,m_3^2 s\) \int_x^y\!{\cal D}z \,
\[\;{\cal P}
\exp\!\(- \int_0^s\!\! d \tau \( \,\frac{1}{4}\,\dot{z}_k^2(\tau) 
\pl ig_3 \bm{A}_k \dot{z}_k \,\) \)\,
\]_{\alpha,\beta}
\;\delta_{ab}
\nonumber
\eea
where $m_3^2$ is the squared mass of the fundamental scalar.
$s$ is the Schwinger proper time. $z_k(\tau)$ is a path which connects
$x$ with $y$ and is parameterized by $\tau$; $\tau$ runs from 0 to $s$.
$\dot{z}_k(\tau)$ denotes the derivative of  $z_k(\tau)$
with respect to $\tau$.

The path integral $\int {\cal D}z$ 
is performed over all paths from $x$ to $y$. 
The measure of the integral is the exponential 
function of the wordline action
\be\label{Weltlinienwirkung}
{\cal P} \exp\!\(- \int_0^s\!\! d \tau \( \,\frac{1}{4}\,\dot{z}_k^2(\tau) 
\pl ig_3 \bm{A}_k \dot{z}_k \,\)\)\\[1ex]
\quad.
\ee
The first term in the exponent alone describes the free propagation
of a scalar field.
The second term includes the interaction with the gauge field $\bm{A}_k\,$;
$g_3$ is the three dimensional gauge coupling.
Scalar self-interactions are neglected (quenched approximation).
The exponential function is ordered along the path; ${\cal P}$ denotes
the path-ordering operator.
The gauge field in \eqn{gfktfs1} is treated as a fixed background field. 
Later on we will quantize averaging over this field. 
(In principle there is one worldline action 
for every index pair $(\alpha,\beta)$.)

The path-ordered exponential function  \eq{Weltlinienwirkung} 
is matrix valued. 
It is identical to 
\bea
&& 
\exp\!\(-\, \frac{1}{4} \int_0^s\!\! d \tau \dot{z}_k^2(\tau) \)
\[\,
{\cal P} \exp\!\(\,-\, ig_3 \int_0^s\!\! d \tau  \bm{A}_k \,\dot{z}_k\, \)
\,\]_{\alpha,\beta} \nonumber\\[1ex]
&=&
\exp\!\(-\, \frac{1}{4} \int_0^s\!\! d \tau \dot{z}_k^2(\tau) \)\,
\bm{T}(x,y)_{\alpha, \beta}
\quad,
\eea
where $\bm{T}(x,y)$ is the link-operator from \eqn{linkop}.

Using this the Green's function \eq{gfktfs1} can be written 
in Feynman-Schwinger-re\-pre\-sen\-ta\-tion 
\be\label{gfktfs}
\bm{G}(x,y,\!\bm{A})_{a \alpha b \beta} \gl
\int_0^\infty \!\!ds \exp\!\(-\,m_3^2 s\) \int_x^y\!{\cal D}z
\exp\!\(-\,\frac{1}{4} \int_0^s \!\! d\tau\, \dot{z}_k^2(\tau)\)\, 
\bm{T}(x,y)_{\alpha\beta}\;\delta_{ab} 
\quad.
\ee
The path used for the evaluation 
of $\bm{T}(x,y)$ is the one integrated over in the path integral. 
All interactions of the scalar with the gauge field are expressed 
by this link-operator.

\paragraph*{The Green's function of a pair of scalars}

The Green's function  $G(x,\bar{x},y,\bar{y})$ connecting 
the two isospin singlet operators
$\Tr \PH(x)^\dagger \bm{T}(x,\bar{x}) \PH(\bar{x})$ and 
$\Tr \PH(y)^\dagger \bm{T}(y,\bar{y}) \PH(\bar{y})$ 
is built up by two fundamental Green's functions \eq{gfktfs} and two link
operators $\bm{T}$
\bea
\mqquad&&
G(x,\bar{x},y,\bar{y}) \label{gfktbs2}\\[1ex]
\mqquad&&=\;\;
\!\Big\langle 
\bm{G}(\bar{x},\bar{y},\!\bm{A})_{a \alpha b \beta}\; 
\bm{T}(\bar{y},y)_{\beta\gamma}\;
\bm{G}(y,x,\!\bm{A})_{b \gamma a \delta} \; 
\bm{T}(x,\bar{x})_{\delta\alpha}
\Big\rangle_{\!\!\bmind{A}}
\nonumber\\[1ex]
\mqquad&&=\;\;2\;
\int_0^\infty \!\!ds \int_0^\infty \!\!d\bar{s} \;
           \exp\( -\,m_3^2(s + \bar{s}) \) \;
             \int\!\! \int \! {\cal D}z\, {\cal D}\bar{z}\,
\nonumber\\
\mqquad&&
\qquad\quad \times\;\; 
 \exp\!\( -\,\frac{1}{4}\int_0^s\!\! d\tau \,\dot{z}_k^2(\tau) 
 \mi\frac{1}{4}\int_0^{\bar{s}}\!\!d\bar{\tau} \,\dot{\bar{z}}_k^2(\bar{\tau}) 
         \)
        \left\langle {\cal P} \exp\( -ig_3 \oint_{x,\bar{x},\bar{y},y} 
         \mquad\bm{A}_k\, dz_k\) \right\rangle_{\!\!\bmind{A}}  \nonumber
\quad.
\eea
The brackets $\langle \ldots \rangle_{\bmind{A}}$ 
symbolize the functional integration over the gauge field $\bm{A}$. 
In principle this averaging contains all perturbative and non-perturbative effects.
The last factor is the expectation value of the  Wegner-Wilson-loop in a
gauge field background. 
It results from four link-operators; two in the 
Green's functions of the fundamental scalars \eq{gfktfs} and two in the 
link-operators of the bound states (eq.~\ref{nlsing}).
The paths  $z(\tau)$ and $\bar z(\bar\tau)$ determine the both sides of
the Wegner-Wilson-loop.

The Green's function of a triplet is correspondingly
\bea
\mqquad&&
G_T(x,\bar{x},y,\bar{y}) \label{gfktbstrip}\\[1ex]
\mqquad&&=\;\;
\!\Big\langle 
\bm{G}(\bar{x},\bar{y},\!\bm{A})_{a \alpha b \beta}\; 
\bm{T}(\bar{y},y)_{\beta\gamma}\; \tau^i_{bc} \;
\bm{G}(y,x,\!\bm{A})_{c \gamma d \delta} \; 
\bm{T}(x,\bar{x})_{\delta\alpha}\; \tau^j_{da} \;
\Big\rangle_{\!\!\bmind{A}}
\nonumber\\[1ex]
\mqquad&&=\;\;
\delta^{ij} \; G(x,\bar{x},y,\bar{y}) 
\eea
For a given \,$i\tgl j$\, the Green's function of a triplet is 
identical to the Green's function of a singlet,
but due to the selection rules discussed in section  \ref{scalarboundstates}
the iso-singlet and iso-triplet states must have different angular momenta.
 
\paragraph*{The correlation mass}

The correlation mass of the bound states $\calM$ describes the fall 
off of the Green's function at large distances $\Theta$
\be
\label{DefKorMass}
G(x,\bar{x},y,\bar{y}) \;\;\propto\;\; \exp(\:\!-\,\calM\:\!\Theta\,)
\quad.
\ee
The extension of the states has to be small as compared to their distance
\be
\label{Theta}
|x-\bar{x}|,\; |y-\bar{y}| \quad\ll\quad |x-y|,\; |\bar{x}-\bar{y}|
\quad \approx \quad \Theta
\quad.
\ee
An exact definition of $\calM$  is given by
\be\label{defcorrmass}
\calM \gl
-\;\lim_{\Theta\ra\infty}\; \fr1\Theta \,\ln\(\, G(x,\bar{x},y,\bar{y})\, \) 
\quad.
\ee
From the form of the Green's function  $G(x,\bar{x},y,\bar{y})$ it follows that 
$\calM$ depends only on $m_3^2$ and on the expectation value of the 
Wegner-Wilson-Loop. 
The other variables used in \eqn{gfktbs2} are integrated out.
All possible angular momenta contribute to $G(x,\bar{x},y,\bar{y})$.
They have to be isolated later.

\section{The calculation of the correlation masses}\label{BerKorrMass}

As in Simonovs original work \cite{Simonov3} 
the correlation masses are calculated from \eqn{defcorrmass}
by simplifying the Green's function of the bound state.
The calculations are, however, reorganized and generalized.
Furthermore we have in our (scalar) case no problems with 
chiral symmetry-breaking and spin-spin-interactions.
In order to gain  a manageable form we have to make two approximations.

\paragraph*{The parameterization of the paths}

In a first step the paths of both scalars are parameterized by a common
parameter $\gamma$ via
\be
\tau \gl \gamma\,s
\qquad\qquad
\bar\tau \gl \gamma\,\bar s
\qquad\qquad{\rm with} \qquad\qquad
0\leq\gamma\leq 1
\quad.
\ee
The kinetic terms of \eqn{gfktbs2} become
\bea
&&
\frac{1}{4}\int_0^s\!\! d\tau \,\dot{z}_k^2(\tau) \pl
             \frac{1}{4}\int_0^{\bar{s}}\!\!d\bar{\tau} \,
             \dot{\bar{z}}_k^2(\bar{\tau})
\nonumber\\[1ex] 
&=&
\frac{1}{4}\int_0^1\!\! d\gamma 
              \(\fr1s \(\fr{\del z_k(\gamma)}{\del\gamma}\)^2 \pl 
              \fr1{\bar s} \(\fr{\del \bar z_k(\gamma)}{\del\gamma}\)^2 \)
\quad.
\label{kinTerme1} 
\eea

The spatial vectors $z_k(\gamma)$ and $\bar z_k(\gamma)$ 
are then replaced by a kind of ``center of mass'' 
and a relative coordinate
\be\label{Sub}
R_k \gl \frac{s \bar{s}}{s\!+\!\bar{s}} 
          \(\, \frac{1}{s} z_k + \frac{1}{\bar{s}} \bar{z}_k\) 
\qquad\qquad\qquad
u_k \gl z_k - \bar{z}_k \quad.
\ee
For $z_k$ and $\bar z_k$ one gets
\be
z_k \gl R_k \pl \fr{s}{s\!+\!\bar s}\;u_k
\qquad\qquad\qquad
\bar z_k \gl R_k \mi \fr{\bar s}{s\!+\!\bar s}\;u_k
\quad.
\ee

The path integrals $\int\!\int {\cal D}z\, {\cal D}\bar{z}$  in  
\eqn{gfktbs2} transform into path integral over the new variables
 $\int\!\int {\cal D}R\, {\cal D}u$.
and the kinetic terms \eq{kinTerme1} become
\be
\frac{1}{4}\int_0^1\!\! d\gamma \(\; 
\fr{s\!+\!\bar s}{s\bar s} \(\fr{\del R_k(\gamma)}{\del\gamma}\)^2  \pl
\fr{1}{s\!+\!\bar s} \(\fr{\del u_k(\gamma)}{\del\gamma}\)^2 \)
\quad.
\ee
Mixed terms cancel. 
This parameterization is always possible and no approximation has been made.

\paragraph*{The classical path of the center of mass}

Now as in \cite{Simonov3} we make the crucial assumption that the 
classical path dominates the trajectory of the center of the bound state. 
This approximation is justified for
heavy bound states. It is not obvious
which scale has to be used as reference,
therefore we can not decide here if the ``large mass condition''
is fulfilled; we will come back to this
question in section \ref{discussion}.

In this approximation the path integral $\int {\cal D}R$ can be replaced by
an appropriate parameterization of $R_k(\gamma)$. 
We choose the $x_3$-axes as the direction of the path
\be\label{Massenzentrumsbewegung}
R_1 \gl R_2 \gl 0  \qquad\qquad\qquad R_3 \gl \gamma\, \Theta \gl \vartheta
\quad,
\ee 
where $\Theta$ is the distance between the two bound states introduced in 
\eqn{Theta}. 
The parameter $\vartheta$ will be interpreted in connection with 
\eqn{quasieuklwirkung}.

The Green's function of the bound state becomes
\bea
\mquad\!\!\!&&
G(x,\bar{x},y,\bar{y}) \label{gfktbs3}\\[1ex]
\mquad\!\!\!&&\propto\;\;
\int_0^\infty \!\!ds \int_0^\infty \!\!d\bar{s}\; 
            \exp\!\( -\,m_3^2(s\!+\!\bar{s}) \) 
             \int \! {\cal D}u 
           \quad \nonumber\\
\mquad\!\!\!&&\times\;\;
  \exp\!\(  - \,\frac{1}{4}\int_0^1\!\!d\gamma \( 
          \fr{s\!+\!\bar s}{s\bar s}\, \Theta^2 \pl
          \fr{1}{s\!+\!\bar s} \(\fr{\del u_k(\gamma)}{\del\gamma}\)^2 \)
         \)
        \left\langle {\cal P} \exp\!\( -\,ig_3 \oint_{x,\bar{x},\bar{y},y} 
         \mquad\bm{A}_k\, dz_k\) \right\rangle_{\!\!\bmind{A}}  
\;. \nonumber
\eea

\paragraph*{The modified area law}

The Wegner-Wilson-Loop is unchanged by the former transformations.
Motivated by lattice results we propose the ansatz 
\be\label{modarealaw}
\left\langle\! {\cal P} \exp\!\( -ig_3 \oint_{x,\bar{x},\bar{y},y} 
         \mquad\bm{A}_k\, dz_k\,\) \right\rangle_{\!\!\bmind{A}}
\;\;\propto\;\;
\exp\( -\; \Theta \int_0^1\!\!d\gamma\; 
        V\!\(\sqrt{u_1^2(\gamma)+u_2^2(\gamma)}\) \)
\quad
\ee
for the gauge field averaged expression.
This replacement generalizes the evaluation of the Wegner-Wilson-loop
in the references \cite{Simonov3} and \cite{Wir3}.
As we will see in section \ref{KorrMassNonlinPot}, 
effects important at small distances $r$ 
can be taken into account (cf.\ sect.~\ref{KorrMassNonlinPot}) with this generalization.

Note that the averaging over the gauge field is included in the 
modified area law \eq{modarealaw}. 
The non-perturbative effects originate in the 
gauge boson sector and are represented by this averaging.
We thus 
assume that all non-perturbative effects on the 
bound state masses can be expressed by a scalar-scalar-potential $V(r)$. 

Eq.~(\ref{modarealaw}) evaluated for rectangular Wegner-Wilson-loops 
with infinite length $T$ leads to the two-dimensional potential
\be\label{DefV}
V(r) \gl 
-\; \lim_{T\ra\infty}\; \fr1T\, \ln( W(r,T) )
\quad.
\ee
Here $W(r,T)$  denotes the expectation value of the rectangular loop
with length $T$ and width $r$. 
In section  \ref{Vlattice} we use this definition to determine $V(r)$ 
from lattice data.

Using the replacement of \eqn{modarealaw} the  $u_3$-path integral
in the Green's function of the bound state  \eq{gfktbs3}
becomes independent of $\Theta$. Therefore it does not contribute to
the correlation mass. Since the Green's function is only needed up
to a constant we disregard this contribution.
The two remaining coordinates are combined to a vector 
\,$\vec u\tgl(u_1, u_2)$\,.

\paragraph*{Evaluation of the Green's function}

In the next step we find a Hamiltonian for a corresponding 
Schr\"odinger type  
equation whose Green's function is the one of \eqn{gfktbs3} with the 
replacement of \eqn{modarealaw}. The latter should bahave as 
\,$G\propto \exp(-M_i\Theta)$\, in the channel $i$.
This has two advantages: First the separation
of states with different angular momentum is very simple and second the
solution of the resulting differential equations is much simpler than
performing the path integrals. 

For this purpose it is advisable to introduce new variables. 
First $\gamma$ is replaced by the parameter $\vartheta$ 
from \eqn{Massenzentrumsbewegung}.
Then the Schwinger proper time variables are substituted by 
\be\label{mus}
\mu \gl \frac{\Theta}{2 s} \qquad \qquad
\bar\mu \gl \frac {\Theta}{2 \bar{s}} \qquad \qquad
\tilde{\mu} \gl  \frac{\mu \bar\mu}{\mu + \bar\mu} \gl  
             \fr{\Theta}{2\,(s\!+\!\bar s)}
\quad.
\ee
The Green's function \eq{gfktbs3} becomes
\bea\label{gfktbs4}
\mquad\!\!\!&&
G(x,\bar{x},y,\bar{y})\\[1ex]
\mquad\!\!\!&& \propto\;\,
\int_0^\infty \!\!d\mu \int_0^\infty \!\!d\bar{\mu}\; 
\(\fr{\Theta}{2\mu\bar\mu}\)^2
\exp\!\( -\,\fr{\Theta}{2} 
      \[\,  m_3^2 \( \fr{1}{\mu} \!+\! \fr{1}{\bar\mu} \) +
    \( \mu\!+\!\bar\mu\) \,\] \)
\int \! {\cal D}\vec{u} \; \exp(-B) \nonumber
\eea
with
\be\label{quasieuklwirkung}
B \gl 
\int_0^\Theta \!\!d\vartheta \[\; 
   \fr{\tilde\mu}{2} \( 
        \frac{\partial \vec{u}(\vartheta)}{\partial \vartheta} \)^2 \pl
   V(|\vec u|) \;\]
\quad.
\ee

If we look first to the $\vec{u}$-path integral the formal analogy to 
a two dimensional system is obvious: a particle with mass $\tilde\mu$
at position $\vec{u}$ and time $\vartheta$ moves in the potential 
$V\(|\vec{u}|\)$. The Euclidean action is $B$; the time interval 
runs from 0 to $\Theta$. Hence, the component \tGl{\vartheta}{R_3} 
of the center of mass coordinate takes over the role of the quasi time.

\paragraph*{The Schr\"odinger equation of the two dimensional problem.}

The time independent Schr\"odinger equation  corresponding to the 
two dimensional problem is
\be\label{Schroedingergl}
H\; \Psi(\vec u) \gl \epsilon \; \Psi(\vec u) 
\ee
where $\epsilon$ is the energy eigenvalue and the Hamilton operator is
given by 
\be
H \gl - \;\frac{1}{2\tilde{\mu}} \( \frac{\partial^2}{\partial u_1^2} 
          + \frac{\partial^2}{\partial u_2^2} \) 
          \pl  V\(|\vec u|\)\quad.
\ee
Due to the rotational symmetry of $H$ we can separate the radial 
from the angular variables and the problem simplifies to the 
solution of the radial equation. Using \,$r\tgl|\vec u|$\, one gets
\be\label{Radgl}
- \,\frac{1}{2\tilde{\mu}}
\( \fr{\del^2}{\del r^2} \pl \fr1r \fr{\del}{\del r} \mi \fr{l^2}{r^2} 
\) \psi_{nl}(r) \pl
V(r)\; \psi_{nl}(r)
\gl
\epsilon_{nl} \; \psi_{nl}(r)
\quad,
\ee
where \,$n\tgl 1,  2, \ldots$\, is the radial and 
\,$l\tgl 0, \pm 1, \pm 2, \ldots$\, is the angular momentum
quantum number. 

If $V(r)$ is less singular than $1/r$ at \tGl{r}{0},
the boundary conditions are 
\be\label{randbed}
\lim_{r\ra\infty}\;\psi_{nl}(r)\gl 0
\qquad \qquad {\rm and} \qquad \qquad
\psi_{nl}(r) \gl r^{|l|} 
\qquad {\rm at} \qquad r\gl0
\quad.
\ee
For a given potential $V(r)$ and fixed quantum numbers  $n$ and $l$ 
it is possible to solve the radial equation and to calculate the 
eigenvalue $\epsilon_{nl}$. In section \ref{KorrMassLinPot} 
we derive the results for some potentials of interest.

\paragraph*{The Green's function of a particular state}

The solution of the Schr\"odinger equation (\ref{Schroedingergl})
allows us to express the $u$-dependent part of the Green's function 
by
\be
\int\! {\cal D}\vec u \; \exp(- B)
\gl
\sum_{n,l}\;\psi_{nl}^*(\vec u(\Theta))\; \psi_{nl}(\vec u(0)) \,
\exp(-\epsilon_{nl} \Theta)
\ee
Thus the contribution of a particular state with quantum numbers 
$n$ and $l$ to the $\Theta$-dependence of the Green's function \eq{gfktbs4}
is given by 
\bea
\mqquad && 
G_{nl}(x,\bar{x},y,\bar{y}) \label{gfktbs5}\\[1ex]
\mqquad && 
\propto\;\;
\int_0^\infty \!\!d\mu \int_0^\infty \!\!d\bar{\mu}\; 
\(\fr{\Theta}{2\mu\bar\mu}\)^2
\exp\!\( \,-\,\fr{\Theta}{2} 
       \[\;  m_3^2 \( \fr{1}{\mu} \!+\! \fr{1}{\bar\mu} \) \pl
    \( \mu\!+\!\bar\mu\) \pl 2\, \epsilon_{nl}(2\tilde\mu)\;\] \)
\quad.
\nonumber
\eea

\paragraph*{The saddle point method}

The $\mu$- and the $\bar\mu$-integrals are evaluated with the saddle point 
method. For large values of $\Theta$ both integrals are dominated 
by the exponential function. In the limiting case of 
\,$\Theta\!\rightarrow\!\infty$\,
only the maximum of the exponent with respect to $\mu$ and $\bar\mu$ 
contributes.

The calculation may by abbreviated
using the fact that the exponent is symmetric in $\mu$ and $\bar\mu$.
This is due to the fact that \eqn{gfktbs2} is symmetric in
$s$ and $\bar s$. This symmetry in turn originates from the identity of
the masses of both fundamental scalars. 
The saddle point equations for $\mu$ and $\bar\mu$ are hence equivalent.
The minimum of both is unique and therefore the same. 
It is possible to identify both parameters. Using 
\,$\mu\tgl\bar\mu\tgl2\tilde\mu$\,  (cf.\ eq.\ \ref{mus})  we
arrive at the Green's function 
\be
G_{nl}(x,\bar{x},y,\bar{y}) 
\;\;\propto\;\;
\exp(\,-\,\calM_{nl}(\mu)\,\Theta\,)
\ee
with the correlation mass 
\be\label{KorrMass}
\calM_{nl}(\mu) \gl \frac{m_3^2}{\mu} \pl \mu \pl \epsilon_{nl}(\mu)
\quad,
\ee
where $\mu$ fulfills the saddle point equation 
\be\label{sattelpktgl}
\frac{\partial \calM_{nl}(\mu)}{\partial \mu} \gl
-\;\frac{m_3^2}{\mu^2} \pl 1 \pl \fr{\del\epsilon_{nl}(\mu)}{\del\mu} \gl
0
\quad.
\ee

Thus the calculation of the correlation mass from the Green's function 
of the bound state \eq{gfktbs2} is reduced to the solution of a differential
equation \eq{Radgl} and a minimization problem \eq{sattelpktgl}.

\section{The correlation masses}\label{KorrMassLinPot}

\subsection{Linear potential}

The solutions of the Schr\"odinger equation \eq{Schroedingergl}
and the corresponding eigenvalues $\epsilon_{nl}$ depend on the 
scalar-scalar-potential.
In this subsection we investigate the simple and important case of a
linear potential
\be\label{Vlin}
V(r) \gl \sigma\,r
\quad.
\ee

\paragraph*{The radial equation}

Substituting $r$ in \eqn{Radgl} by the dimensionless variable
\be\label{scaleing}
\rho\gl(\mu\sigma)^{1/3}r
\ee 
results in the eigenvalue equation
\be\label{Radglscaled}
- \;\fr{\del^2\psi_{nl}(\rho)}{\del\rho^2} \mi\frac{1}{\rho}\fr{\del\psi_{nl}(\rho)}{\del\rho} 
        \pl \(\frac{l^2}{\rho^2} + \rho \)\psi_{nl}(\rho)
\gl a_{nl}\, \psi_{nl}(\rho) 
\quad.
\ee
The boundary conditions \eq{randbed} are unchanged.

\begin{figure}[t]
\begin{picture}(14.0,7.5)
\put(1.1,5.9){$\psi_{nl}(\rho)$}
\put(1.2,-0.2){\epsfxsize12cm \epsffile{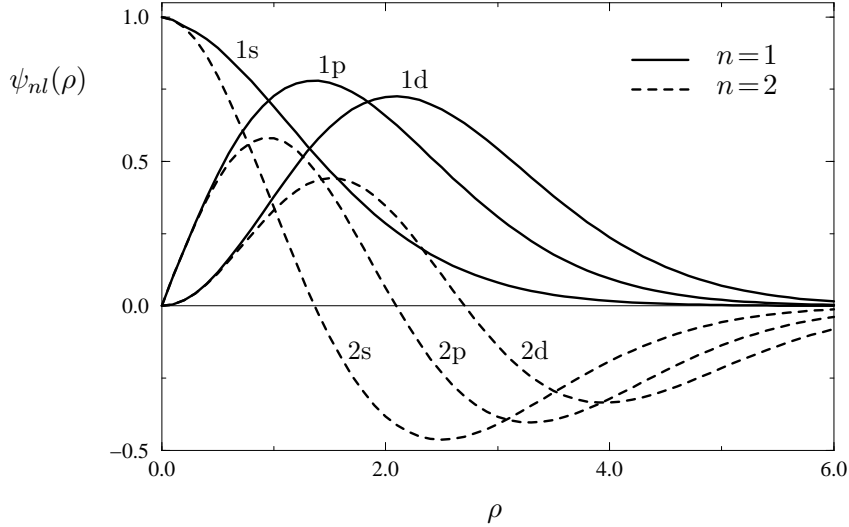}}
\put(10.5,6.22){$n\tgl1$}
\put(10.5,5.82){$n\tgl2$}
\put(7.45,0.2){$\rho$}
\put(4.1,6.3){\small 1s}
\put(5.2,6.1){\small 1p}
\put(6.3,5.9){\small 1d}
\put(5.6,2.3){\small 2s}
\put(6.8,2.3){\small 2p}
\put(7.9,2.3){\small 2d}
\end{picture}
\caption{The solutions of the radial equation (\protect\ref{Radglscaled})
for the quantum numbers \,$n\tgl1,\,2$\, and \,$l\tgl0,\,1,\,2$.
The Green's function of the bound states  (\protect\ref{gfktbs4})
can be expressed be these solutions.
The notation  s-, p- and d-wave corresponds to different angular
momentum quantum numbers $l$, as in atomic physics.}
\label{AbbSolsRadEq}
\end{figure}

The numerical solution of this differential equation does not cause 
any difficulties. The eigenfunction with the lowest 
quantum numbers are displayed in figure \ref{AbbSolsRadEq}. 
The corresponding eigenvalues  $a_{nl}$ are:
\begin{center}
\begin{tabular}{|r|lll|}
\hline
$a_{nl}$ & $l$=0 & $l$=1 &  $l$=2 \\
\hline
$n$=1 & 1.74  & 2.87  &  3.82 \\
$n$=2 & 3.67  & 4.49  &  5.26 \\
\hline
\end{tabular}
\end{center}

The eigenvalues of interest  $\epsilon_{nl}(\mu)$ can by calculated
from the dimensionless eigenvalues $a_{nl}$ of \eqn{Radglscaled} via
\be\label{epsilonlinPot}
\epsilon_{nl}(\mu) \gl \frac{\sigma^{2/3}}{\mu^{1/3}} \;a_{nl}
\quad.
\ee
Therefore the $\mu$-dependence of $\epsilon_{nl}(\mu)$ is a power law
in the case of a linear potential.

\paragraph*{The saddle point equation}

Using $\epsilon_{nl}(\mu)$ from \eqn{epsilonlinPot} the saddle point 
equation (\ref{sattelpktgl}) becomes
\be\label{sattelpktgllinPot}
-\;\frac{m_3^2}{\mu^2} \pl 1 \mi \fr13 \, \mu^{-4/3} \sigma^{2/3} a_{nl} \gl
0
\quad.
\ee
It is solved by 
\begin{equation}\label{mu}
\mu \gl  z\!\(a_{nl}, b\)^{3/2}\,\sqrt{\sigma}
\qquad \qquad {\rm with} \qquad \qquad
b \gl \frac{m_3^2}{\sigma}
\end{equation}
where $z(a,b)$ is a positive and real solution of the cubic equation 
\be
z^3 \mi \frac{1}{3} a\, z \mi b \gl 0
\quad.
\ee
The solution corresponding to the minimum of $\calM(\mu)$ is
\be\label{solz}
z(a,b) \gl \frac{2^{1/3}}{3} a \( 27 b + \sqrt{729 b^2 - 4 a^3}\)^{-1/3}
+\; \frac{1}{2^{1/3}\,3}\( 27 b + \sqrt{729 b^2 - 4 a^3}\)^{1/3}
\quad.
\ee
It is unique if \,$b\!\geq\!0$\,; for 
\,$-\tfr{2}{27}a^{3/2}\!<\!b\!<\!0$\,
it corresponds to a local minimum of $\calM$.
If $b$ is even smaller there is no positive and real solution anymore.
We therefore treat only the case that $m_3^2$ and hence $b$ is positive.

\begin{figure}[t]
\begin{picture}(14.0,7.5)
\put(1.0,6.2){$\nfr{\calM_{nl}}{\sqrt{\sigma}}$}
\put(1.2,-0.2){\epsfxsize12cm \epsffile{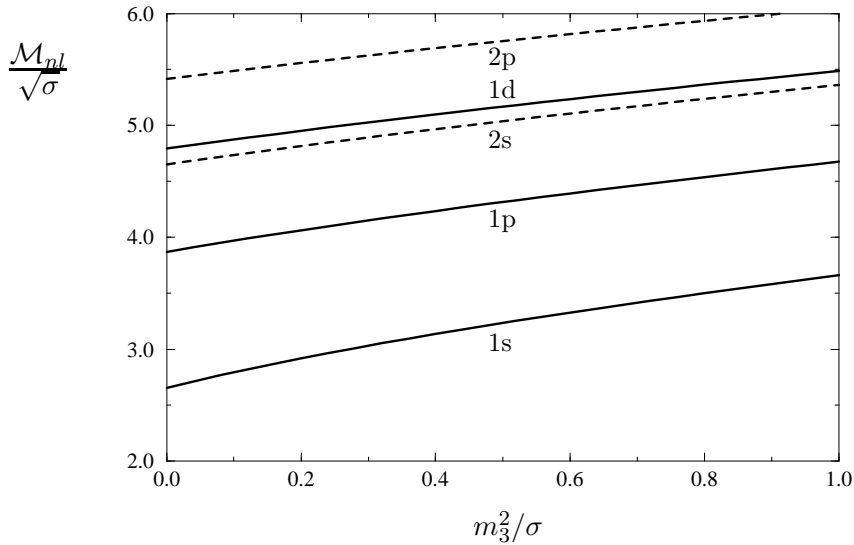}}
\put(7.2,0.1){$m_3^2/\sigma$}
\put(7.4,6.36){\small 2p}
\put(7.4,5.92){\small 1d}
\put(7.4,5.25){\small 2s}
\put(7.4,4.2){\small 1p}
\put(7.4,2.55){\small 1s}
\end{picture}
\caption{The correlation mass $\calM_{nl}$ of the bound states 
versus the squared mass of the fundamental scalar $m_3^2$ for a linear 
potential. We find a spectrum of higher excitations. 
The string tension $\sigma$ sets the scale of the bound state masses. 
In the case of 
a linear scalar-scalar-potential the dependence of the bound state
mass on the mass of the fundamental scalar can be given analytically.}
\label{AbbTermschema}
\end{figure}

The correlation mass \eq{KorrMass} evaluates to
\be
\frac{\calM_{nl}}{\sqrt{\sigma}} \gl 
4\, z\!\(a_{nl} , \frac{m_3^2}{\sigma}\)^{3/2} \mi 
2\, z\!\(a_{nl} , \frac{m_3^2}{\sigma}\)^{-3/2} \frac{m_3^2}{\sigma} 
\quad,
\ee
with the function $z$ given in \eqn{solz}.
It depends only on the dimensionless eigenvalue $a_{nl}$, on the 
string tension $\sigma$ and on the squared mass $m_3^2$ of the fundamental 
scalar. The result scales with $\sigma$.
In figure \ref{AbbTermschema} we plot the correlation masses of the 
bound states with the lowest quantum numbers versus the squared mass
of the fundamental scalar. 

Our results show a hierarchy of bound states. The lowest state 
corresponds to the 1s-wave ($n\tgl1$, $l\tgl0$).
Due to the selection rules derived in section \ref{scalarboundstates} 
it has to be an isospin-singlet. The 1p-state 
($n\tgl1$, $l\tgl1$) is heavier, it is a vector-isospin-triplet.
A dense spectrum of higher excitations follows. 
Note that the string tension determines the typical scale of the bound
state masses. It is a genuine non-perturbative quantity.

For the case that \,$m_3^2 \lsim \sigma$\, our treatment is by no means an 
analogue of the non-relativistic quark model . This can be seen by comparing \eqn{Radgl}
with \eqn{KorrMass}. From (\ref{KorrMass}) we would deduce a constituent mass
of the elementary scalars
\be
m_{\rm constit} \gl \fr{\mu}{2} \pl \fr{m_3^2}{2\mu}
\ee
whereas from the kinetic terms we would deduce \tGl{m_{\rm constit}}{\mu}.
For \,$m_3^2\gsim \sigma$\, there is no contradiction since $\mu$
goes to $m_3^2$ in that case. Furthermore the constituent mass 
shows a marked dependence on the state for \,$m_3^2\lsim \sigma$\,.

The mass parameter $\mu$ varies from 0.24$g_3^2$ to 0.4$g_3^2$ going
from 1s to 2p corresponding to a quasi static constituent mass of half that
value at \tGl{m_3^2}{0}.\footnote{Thus it is about half of the constituent mass
introduced in a very recent paper by Buchm\"uller and Philipsen \cite{BuPh2}. 
Note that we also have a binding energy $\epsilon_{nl}$.
The constituent W-boson in  \cite{BuPh2} does not enter directly the mass formula in our 
dynamical approach. However it corresponds to the inverse correlation length
of the gauge field strength and hence determines the binding energy via the
string tension (eq.\ (\ref{stringtension}) and appendix \ref{msv}).}

\subsection{A nonlinear potential}
\label{KorrMassNonlinPot}

The procedure presented above can be extended in a straight
forward way to a general potential \cite{Dis}. For the problem under
consideration several modifications of the linear potential are expected
to occur.

\begin{figure}[t]
\begin{picture}(14.0,5.5)
\put(0.0,5.0){a)}
\put(5.13,5.0){b)}
\put(10.25,5.0){c)}
\put(0.0,3.73){\small $V(r)$}
\put(5.13,3.73){\small $V(r)$}
\put(10.25,3.73){\small $V(r)$}
\put(3.8,1.1){\small $r$}
\put(8.93,1.1){\small $r$}
\put(14.03,1.1){\small $r$}
\put(0.2,2.75){\small $V_{\rm sc}$}
\put(0.2,2.17){\small $V_{\rm vp}$}
\put(5.45,0.58){\small $V_0$}
\put(0.5,1.27){\small $0$}
\put(5.6,1.27){\small $0$}
\put(10.71,1.27){\small $0$}
\put(0.4,-0.9){\epsfxsize14.6cm \epsffile{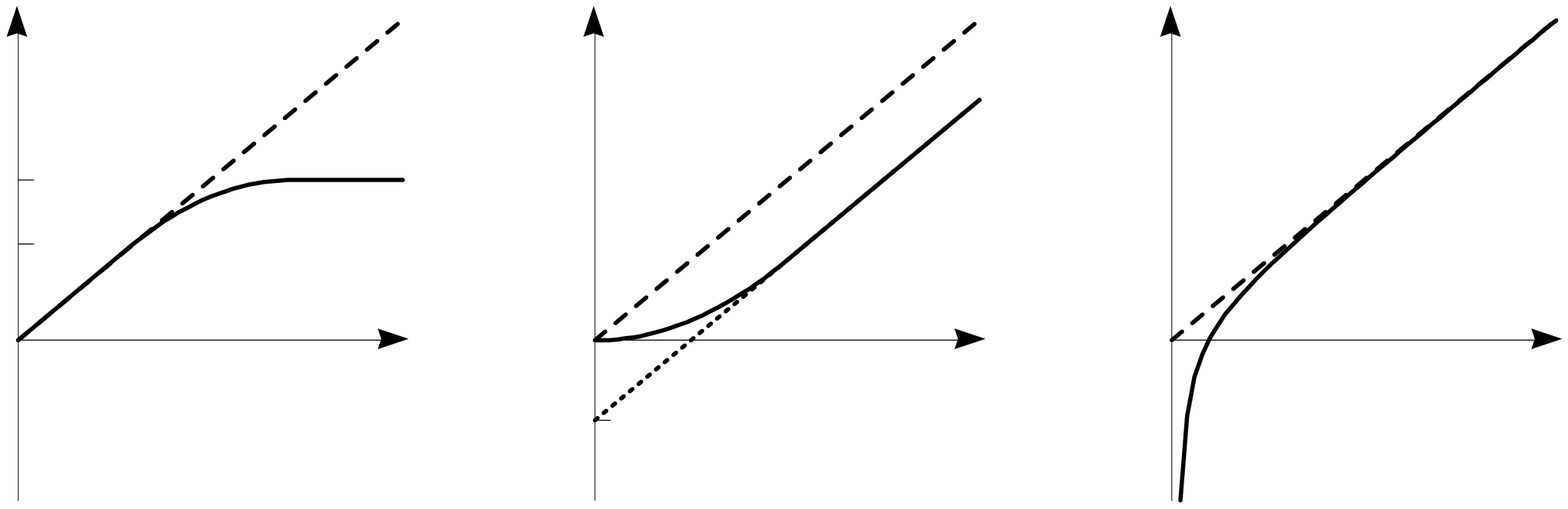}}
\end{picture}
\caption{Different modifications of the linear potential.
a) the influence of screening,
b) the lowering of the potential due to non-perturbative effects,
c) a perturbative contribution.}
\label{AbbModlinPot}
\end{figure}

As in QCD we do expect screening of the potential by spontaneous creation of 
a pair of two fundamental scalars.
The potential would look like the full line in figure \ref{AbbModlinPot}a.
The dashed line shows the corresponding linear potential.
So far lattice results do not indicate any screening and the
low lying bound states should not be affected by the screening anyway. 
We therefore neglect this effect. 

From lattice calculations the potential can only be determined 
up to an additional constant. In the stochastic  
vacuum model, outlined in appendix \ref{msv}, the linear rise of the 
potential sets in only at distances $r$ which are larger than 
the correlation length of two gauge fields in the vacuum. 
The shape of the potential is given in figure \ref{AbbModlinPot}b (full line). 
If the wave functions obtained from the Schr\"odinger equation  
are not dominated by the region of small $r$, the potential  
can be approximated by a linear potential with a negative intercept
\be\label{Vlowerd}
V(r) \gl V_0 \pl \sigma \,r 
\qquad\qquad{\rm with}\qquad\qquad
V_0 \;\;<\;\;0
\ee
(dotted line). The size of the intercept will be discussed in 
section \ref{intercept}.

It is easy to see that the correlation masses corresponding to 
this potential are
\be
\calM_{nl} \gl \calM_{nl}^{\rm lin} \pl V_0
\quad, 
\ee
where $\calM_{nl}^{\rm lin}$ are the masses calculated from the linear
potential. Due to the lowering of the potential the
correlation masses of the bound states are lowered as well.

The difference between the real potential and the 
approximation \eq{Vlowerd} may, if it is small, be taken into account 
as a perturbation (see below).

At small distances $r$ the potential should be
dominated by the  perturbative gauge boson exchange.
The lowest order is calculated in appendix \ref{pertcont}.
This leads to the two dimensional Coulomb potential
\be
V_{\rm C}(|\vec u|)
\gl
\fr{3}{8\pi}\, g_3^2 \,\ln(\,\Lambda\, |\vec u|\,) 
\quad.
\ee
The constant $\Lambda$ cannot be fixed due to the logarithmic
divergence.

If the exchanged particle acquires the mass  $m$ we obtain the 
two dimensional Yukawa potential
\be
V_{\rm Y}(|\vec u|) 
\gl
-\; \fr{3}{8\pi}\,g_3^2\, K_0(\,m\,|\vec u|\,) 
\quad,
\ee
where $K_0$ is the modified Bessel function.
Note that both potentials merge at small distances
\,$|\vec u|\tgl r\!\ll\! m^{-1}$\, with a suitable choice of $\Lambda$.
At these distances the mass does not make any difference. 

If the perturbative contribution to the potential vanishes in the limit 
\,$r\!\ra\!\infty$\, as $V_{\rm Y}(r)$ does, 
the linear potential is not changed here. This situation is 
sketched in figure \ref{AbbModlinPot}c.
It turns out that  
in this case the corrections to the correlation masses $\calM_{nl}$ can be calculated 
perturbatively. One gets
\be\label{Mqmstr}
\calM_{nl} \gl \calM_{nl}^{\rm lin} \pl \delta \calM_{nl}  
\ee
with
\be
\delta \calM_{nl}  \gl
2\pi \int_0^\infty\!\! dr\,r \; V_{\rm Y}(r)\,  \psi_{nl}(r)^2
\quad.
\ee
Since the Yukawa potential is negative, the masses are lowered.

\section{The potential on the lattice}\label{Vlattice}

In order to compare the correlation masses of the bound states with 
lattice data, as will be done in the next section, 
we have to  determine $V(r)$. 
In this section we analyze lattice data of Wegner-Wilson-loops
and fix the parameters of the potential using a suitable fit.

The symmetric electroweak phase of the three dimensional SU(2)-Higgs-model
has been investigated by several groups on the lattice 
\cite{IlgenfEA1,GuertlerEA1,GuertlerEA2,PhilipsenEA,KarschEA,KajantieEA2}.
Nevertheless, only the authors of reference \cite{IlgenfEA1,GuertlerEA1,GuertlerEA2}  
calculated the Wegner-Wilson-loops. 
The data have been reanalyzed by us \cite{Dis} and will be given in a 
form which is appropriate for the purpose of calculating the bound state 
masses.

The lattice data are expressed in terms of  the parameters  
$\beta_G$, $\beta_H$ and $\beta_R$.
For $\beta_G$ there is the simple relation 
\be
\beta_G\gl \fr{4}{a_L\,g_3^2}
\quad,
\ee
where $a_L$ is the lattice constant.
The data we use are evaluated at  \,$\beta_G\tgl 12$\, and hence 
correspond to \tGl{a_L}{\tfr13\, g_3^{-2}}.
The more complicated relation between the other lattice parameters
and the (normalized) continuum parameters \,$\mdrsq\tgl m_3^2/g_3^4$\, 
and \,$\ladr\tgl\ladr/g_3^2$\, \eq{contpar}
are given in reference \cite{Mikko1}.
The data of \cite{IlgenfEA1} correspond to \,$\ladr\tgl 0.0239$\, and 
\tGl{\mdrsq}{0.73};
the \,$\ladr\tgl 0.0957$\, results \cite{GuertlerEA1,GuertlerEA2} 
cover the range \tGl{\mdrsq}{-0.022\;\;{\rm to}\;\;0.524}.

The Wegner-Wilson-loops on the lattice are rectangular and have extensions from
2 to  \,$N\tgl15$\,resp.\ 24 lattice units.
The extrapolation to infinite length is very secure, since the
relevant quantity \,$-\ln W(r,T)$ (cf.\ eq.\ \ref{DefV})
rises practically  linear in $T$ for \,$T\,\gsim\, 8a$.
We found that the lattice data can be reproduced by the  
three parameter fit of the form 
\be\label{FitK}
V(r) \gl
C \mi  \fr{3}{8\,\pi} \,g_3^2\, K_0( m\, r ) \pl \sigma \, r
\ee
The fitted constant $C$ is without any physical significance since it 
depends on the lattice renormalization procedure. 
An effective intercept \tGl{C}{V_0} like in \eqn{Vlowerd} due to non-perturbative effects
will be discussed in section \ref{intercept}.
In table \ref{Fittabelle} we give the results for the 
relevant parameters $m$ and $\sigma$ for the available values  
of $\ladr$ and $\mdrsq$.

\begin{table}[t]
\begin{center}
\begin{tabular}{|c|c|c||c|c|c|}
\hline  \rule[-1.5ex]{0em}{4.5ex}
$\ladr$ & $\mdrsq$ & $N$ & 
$m/g_3^2$ & $\sigma/g_3^4$ \\
\hline \rule[0ex]{0em}{3.2ex}
  0.0957 & ~0.5254 & 15 
& 1.045 & 0.1370  \\
  0.0957 & ~0.3699 & 15 
& 1.105 & 0.1390 \\
  0.0957 & ~0.2153 & 15 
& 1.029 & 0.1345 \\
  0.0957 & ~0.1229 & 15 
& 1.083 & 0.1364 \\
  0.0957 & ~0.0615 & 15 
& 1.046 & 0.1335 \\
  0.0957 & ~0.0309 & 15 
& 0.999 & 0.1296 \\
  0.0957 & ~0.0003 & 15 
& 0.886 & 0.1208 \\
  0.0957 & -0.0058 & 24 
& 0.770 & 0.1135  \\
  0.0957 & -0.0220 & 24 
& 0.411 & 0.0876  \\
  0.0957 & 0.073 &  
& 1.058 & 0.1345  \\
  0.0239 & ~0.073~ & 15 
& 1.051 & 0.1326 \\[0.6ex]
\hline
\end{tabular}
\end{center}
\caption{The fitted parameter of the function (eq.~\protect\ref{FitK}).
The values at \,$\ladr\tgl0.0957$\, and \,$\mdrsq\tgl0.073$\,
are interpolated. $N$ is the number of lattice points, $m$ is the 
effective gauge boson mass and $\sigma$ is the string tension.
}
\label{Fittabelle}
\end{table}

The contribution \,$\tfr{3}{8\pi} g_3^2 K_0(mr)$\, indicates that the
coloured objects exchanged between the scalars have an effective mass
of about $1g_3^2$. It is suggestive to identify it with some  
magnetic mass
of the gauge boson. The mass $m$ of our fit is, however, larger than
the magnetic mass obtained in lattice simulations 
in Landau gauge \cite{KarschEA} and also larger than the one predicted in 
reference \cite{BuPh} from gap equations. 

Comparing 
the potential parameters at \tGl{\ladr}{0.0239} with the ones at 
\tGl{\ladr}{0.0957} (interpolated) one sees practically no effect
of the quartic coupling. This can be understood from the fact
that the exchange of a scalar is suppressed by a factor $\ladr^2$
in comparison to the gauge boson exchange. 

The shape of the scalar-scalar-potential depends at \,$\mdrsq\!\geq\!0.0615$\,
only marginally on $\mdrsq$. In this mass range $V(r)$ is nearly 
exclusively determined by the gauge boson sector.
This behavior is expected for large scalar masses, since the scalars decouple
here. 
From the data of table \ref{Fittabelle} we conclude that the mass-limit
from which on the influence of the scalars on  $V(r)$ is negligible is small
compared to the other mass scales of the problem. 
Above this limit the string tension $\sigma$ should be identical to
the one of the pure SU(2)-Yang-Mills-Theory. 
Indeed, Teper \cite{Teper} finds for the latter one roughly 
$\sigma\approx 0.137g_3^4$  at \tGl{\beta_G}{12} 
based on the correlation of Polyakov-loops.
This value is similar to the string tensions of tabel~\ref{Fittabelle}.  
He extrapolates his results to \tGl{\beta_G}{\infty} in order to gain 
a continuum value of the strig tension. We stay, however, with the 
potential at \tGl{\beta_G}{12}, since we compare with lattice data
at the same value in the next chapter.

The data point \,$\mdrsq\tgl-0.022$\, is at the critical temperature.
The fact that the critical $\mdrsq$ is negative is in agreement with 
other lattice investigations \cite{KajantieEA2}. 

\section{Comparison of the correlation masses with lattice calculations}
\label{VerglGitter}

\paragraph*{The correlation masses in our model}

We calculated the correlation masses $\calM_{nl}$ of the bound states 
with the method presented above. We based this calculations on the potential
of \eqn{FitK} with $m$ and $\sigma$ from table \ref{Fittabelle} and 
the arbitrary choise \tGl{C}{0}. 
The effect of the term \,$-\tfr{3}{8\pi} g_3^2 K_0(mr)$\, is very small.
It lowers the masses by less than 2\% and can safely be treated 
perturbatively. This is due to the large extensions of the bound states as
compared to $1/m$. 
Our results for the correlation masses are listed in 
table \ref{Massentabelle}.

\begin{table}
\begin{center}
\begin{tabular}{|c|l||c|c|c|c|c|}
\hline  
$\ladr$ & $\quad\mdrsq$ & 
$ \tfr{\calM_{\rm 1s\rule[-0.24ex]{0em}{0.24ex}}}{g_3^2} $ & 
$ \tfr{\calM_{\rm 2s\rule[-0.24ex]{0em}{0.24ex}}}{g_3^2} $ & 
$ \tfr{\calM_{\rm 1p\rule[-0.24ex]{0em}{0.24ex}}}{g_3^2} $ & 
$ \tfr{\calM_{\rm 2p\rule[-0.24ex]{0em}{0.24ex}}}{g_3^2} $ & 
$ \tfr{\calM_{\rm 1d\rule[-0.24ex]{0em}{0.24ex}}}{g_3^2} $ \rule[-2.2ex]{0em}{5.9ex}\\
\hline
0.0957  & ~0.5254   &   1.93   &   2.48   &   2.27   &   2.71   &   2.53   \\
0.0957  & ~0.3699   &   1.73   &   2.31   &   2.09   &   2.55   &   2.36   \\
0.0957  & ~0.2153   &   1.48   &   2.08   &   1.85   &   2.33   &   2.14   \\
0.0957  & ~0.1229   &   1.31   &   1.95   &   1.70   &   2.21   &   2.00   \\
0.0957  & ~0.0615   &   1.15   &   1.82   &   1.56   &   2.09   &   1.88   \\
0.0957  & ~0.0309   &   1.05   &   1.74   &   1.47   &   2.01   &   1.79   \\
0.0957  & ~0.0003   &   0.91   &   1.61   &   1.34   &   1.88   &   1.67   \\
0.0239  & ~0.073    &   1.18   &   1.84   &   1.58   &   2.10   &   1.89   \\
0.0239  & ~0.089    &   1.22   &   1.87   &   1.62   &   2.13   &   1.92  \\
\hline
\end{tabular}
\end{center}
\caption{The masses of the bound states up to an arbitrary additional 
constant $C$.  They have been calculated using 
the potential eq.~(\protect\ref{FitK}) with the parameters of 
table \protect\ref{Fittabelle} and \tGl{C}{0}.}
\label{Massentabelle}
\end{table}

In lattice calculations the correlations of pairs of the operators 
\eq{nlsing} respectively \eq{nltrip} are investigated. The relative orientation
can be varied. Using special combinations of these pairs it is possible
to extract single angular momentum states. The method of blocking
operators and the diagonalization of mass matrices are used to suppress
the mixing of different states. In this way one gets the correlations 
masses corresponding to  definite spin and isospin quantum numbers. 

The quantum numbers of the 1s-state are the same as those of the 
Higgs-particle; the quantum numbers of the 1p-state are identical with 
those of the $W$-boson. Therefore the operators used to investigate the 
1s- and the 1p-state are the same operators which are used in the broken
phase to investigate the Higgs- and $W$-boson, respectively.
Therefore the 1s-mass is sometimes called Higgs-mass, the 1p-mass is
called $W$-mass in the literature.

Philipsen, Teper and Wittig \cite{PhilipsenEA} calculated the 
correlation masses of the bound states at \tGl{\ladr}{0.0239} 
and \tGl{\mdrsq}{0.089} on the lattice. 
They find a spectrum of higher excitations. The masses of 
the low lying states are determined with high accuracy 
\begin{eqnarray*}
\calM_{\rm 1s}^{\rm PTW} &=& ( 0.839\pm 0.015 )\, g_3^2 \quad,\\ 
\calM_{\rm 1p}^{\rm PTW} &=& ( 1.27\pm 0.06 )\, g_3^2 \quad,\\ 
\calM_{\rm 2s}^{\rm PTW} &=& ( 1.47\pm 0.04 )\, g_3^2 
\quad.
\end{eqnarray*}

Since the Wegner-Wilson-loops have not been measured by Philipsen et al.,
we have to use the potentials analyzed above to compare with our results.
We do not have any data at the values of  $\ladr$ 
and $\mdrsq$ used in reference \cite{PhilipsenEA}. In view of the
small $\mdrsq$-dependence of the potential (cf.\ tab.~\ref{Fittabelle})
we use instead the potential corresponding to \tGl{\ladr}{0.0239} 
and \tGl{\mdrsq}{0.073} to calculate the correlation masses. 
The results are given in the last line of table \ref{Massentabelle}.

Since unfortunately the constant $C$ of the potential cannot be deduced from the 
lattice data, we cannot compare the masses directly, but only the 
mass differences, as given in the table 
\vspace*{0.5em} 
\begin{center}
\begin{tabular}{|c|c|c|}
\hline \rule[-1.3ex]{0em}{4.0ex}
& $(\calM_{\rm 1p}-\calM_{\rm 1s})\, g_3^{-2}$ 
& $(\calM_{\rm 2s}-\calM_{\rm 1s})\, g_3^{-2}$ \\
\hline
\rule{0em}{2.8ex}
our model      & 0.40  &  0.65  \\
Philipsen et al. & $0.43\pm 0.06$  & $0.63\pm 0.05$\rule[-0.8ex]{0em}{0.7ex}\\
\hline
\end{tabular}
\end{center}
Within the errors the data predicted by our model agree very well 
with those from the lattice. 

\begin{figure}[t]
\begin{picture}(14.0,7.5)
\put(1.0,6.2){$\nfr{\calM_{\rm 1s\rule[-0.24ex]{0em}{0.24ex}}
              - \calM^0_{\rm 1s\rule[-0.24ex]{0em}{0.24ex}}}{g_3^2}$}
\put(2.2,-0.2){\epsfxsize12cm \epsffile{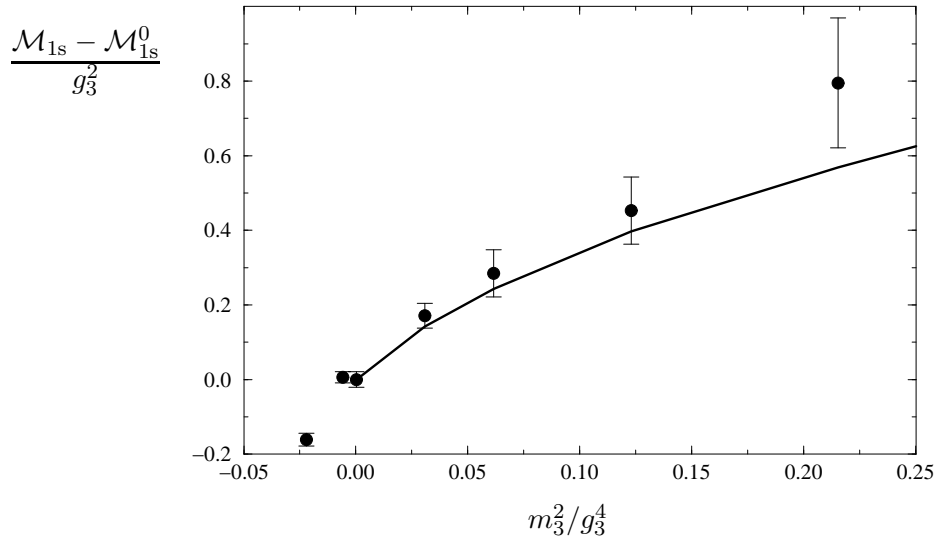}}
\put(7.9,0.1){$m_3^2/g_3^4$}
\end{picture}
\caption{The correlation mass of the 1s-state vs.\ the squared mass of the 
fundamental scalar. The points with error-bars are the lattice results of 
reference \protect\cite{GuertlerEA2}.The full line connects the results of our 
model. Both data sets are normalized at \tGl{m_3^2}{0}.}
\label{AbbMassesvsmdrsq}
\end{figure}

G\"urtler et al.\ \cite{GuertlerEA2} have calculated 
the 1s-masses as well as an upper bound on the 1p-masses at \tGl{\ladr}{0.0957}.
There results for the 1s-masses for different scalar masses $\mdrsq$
are plotted in figure \ref{AbbMassesvsmdrsq} in comparison with our results.
Both data sets are normalized at \tGl{m_3^2}{0}. 
One sees that the dependence of the bound state mass on the mass of the 
fundamental scalar is well described by our model. The main contribution
to the variation of $\calM_{\rm 1s}$ is due to the explicit 
$m_3^2$-dependence of the Green's function \eq{gfktbs2},
while the modification of the potential with $m_3^2$ is not significant.

\section{The intercept}\label{intercept}

The constant $C$ in \eqn{FitK} can not be fixed from lattice data of the 
Wegner-Wilson-Loop. One could choose it to give good agreement with the 
results of reference \cite{GuertlerEA2}, resulting in \tGl{C}{-0.44\, g_3^2}.
Similarly one could choose the constant to give the best possible 
agreement with the masses from 
reference \cite{PhilipsenEA}. This would lead to the slightly
larger value \tGl{C}{-0.38\, g_3^2}.
The difference between these two values can in our opinion not
be attributed to the different values of the quartic coupling.
As we showed in section \ref{Vlattice} the latter has only a small influence
on the potential. It is rather due to a small discrepancy 
between the two lattice investigations, which is, however, within the 
statistical error. 

In order to get some theoretical understanding of the origin of the intercept
it seems to be necessary to investigate the vacuum structure of the 
theory.
In the bound state model  we have considered the binding forces between
the fundamental scalars. The parameters were taken from lattice 
calculations of the
Wegner-Wilson loops. The resulting spectra strongly support a picture
of the effective 3-dimensional electroweak theory above the critical
temperature which is very similar to QCD with light scalar quarks, the linear
confinement playing an essential role. Though there is no analytic
proof for confinement in (4-dimensional) QCD there exists a simple
model which yields confinement for non-Abelian gauge theories in a very
natural way, the model of the stochastic vacuum \cite{Dosch1,DoschSi}. In
appendix \ref{msv} we give the essential features of this model for a
three-dimensional theory. The model yields an asymptotically linear
potential and relates the vacuum expectation value of the gauge boson
fields $\langle\, g_3^2 F F \,\rangle$ and their correlation lengths 
$a$ to the string tension. 
For small inter-quark separations the potential is quadratic
and becomes linear only at a distance of a few correlation lengths $a$
and thus corresponds to the situation depicted in figure 3b. If we
denote by $D(z^2)$ the (rotationally invariant)  correlation function
of the gluon fields in the vacuum (see \eqn{FxFy}) the string tension 
is given by:
\be\label{stringtension} 
\sigma \gl \frac{\pi \langle\, g^2 FF \,\rangle}{12 N_C} 
         \int_0^\infty\!\! dz \,z\, D(z^2)
\ee
whereas the effective intercept \tGl{C}{V_0} (see figure 3b) is given by:
\be\label{interceptmsv}
V_0 \gl -\; \frac{\langle\, g^2 FF \,\rangle}{6 N_C} 
         \int_0^\infty\!\! dz \,z^2\, D(z^2)
\ee
From these equations we can deduce:
\be
V_0 \gl -\,K \sigma a
\ee
where $a$ is defined by 
\be
a \gl 
         \int_0^\infty\!\! dz \, D(z^2)
\ee
and $K$ is of the order 1, the numerical value depends on the form of
$D(z^2)$.
Unfortunately 
neither the condensate $\langle\, g_3^2 \,FF \,\rangle$ nor 
the correlation length have been calculated on the lattice so far, 
so we apply here a simplifying 
argument in order to get an idea of the order of magnitude
of $V_0$.

If $a$ is the correlation length between two field strengths $F$ we
expect the correlation between the product of two field strengths $F^2$
to be of order $a/2$.\footnote{Taking $a/4$ (corresponding to 4 gluons 
in the glueball \cite{BuPh2})
would improve the agreement discussed below.
We thank O.~Philipsen for a discussion of this point.} 
Since the product $F^2$ interpolates a glue-ball
($W$-ball)
we expect $2/a$ to be near the glue-ball ($W$-ball) mass. 
In QCD this seems to be
fulfilled qualitatively, with \,$2/a \!\approx\! 1.5$ GeV. The 
$W$-ball mass has been obtained in lattice calculations 
\cite{PhilipsenEA} and found to be \,$m_G\!=\! 1.60\pm0.04   g_3^2$\,. 
Assuming the correlation length to be \,$a\!=\!2/m_G$\,
we obtain together with the string tension of the lattice calculations
(cf.\ tab.\ \ref{Fittabelle})
the value \,$V_0\!=\! - 0.17 K\, g_3^2 $\, which is indeed of the 
same order of magnitude as the constant $C$ 
which we need to fix the absolute 
values of the bound-state masses. It would be extremely
interesting to determine the correlator $\langle\, F(x) T(x,0) F(0) \,\rangle$  
on the lattice in
order to see if 
considerations similar as in QCD lead to confinement in this
3-dimensional theory. 

\section{Discussion and conclusions}\label{discussion}

The comparison  with the lattice data in the
last section shows that the 
correlation masses of physical (gauge invariant) objects 
in the hot electroweak phase
are described well by the proposed model.
It turns out that it is similar to QCD but without the problem of 
chiral symmetry breaking and spin-spin-interactions.
The fact that the dense spectrum of higher excitations predicted by us in 
reference \cite{Wir3} has been confirmed by lattice calculation is 
in our opinion a success of the model on its own.
The numerical agreement of the mass-differences is much better than it could be
expected in view of the approximations made.
Even the $m_3^2$-dependence of the 1s-mass is explained by the model.
In consideration of this fact we will discuss the assumptions 
of our calculations again.

The Green's function of the fundamental scalar \eq{gfktfs1} does not take into
account the interactions with other scalars (quenched approximation).
Nevertheless, the influence of the dynamical scalars is taken into account. 
The fact that the scalar-scalar-potential measured on the lattice
is not the same for all data sets is due the dynamical scalars. 
The $\ladr$-dependence is small, as far as we can conclude this from 
the existing data. The modification of the potential as function of 
the mass $m_3^2$ is only large  
for very small values of this mass. 
The direct perturbative exchange of a scalar can be taken into
account like the perturbative gauge boson contribution. It is, however, 
suppressed by a factor $\ladr^2$ compared to the latter one. 
The scalar contributions to the scalar self-energy is neglected 
by the quenched approximation as well. All these terms are suppressed
by at least a factor $\ladr^2$ when compared to those contributions taken
into account. This approximation is therefore well-founded. 
Only the influence of the dynamical scalars of the scalar-scalar-potential
for very small squared masses $m_3^2$
is of numerical relevance; this effect is included in our calculations.

In the calculation of the correlation masses from the 
Green's function \eq{gfktbs2} in section  \ref{BerKorrMass}
we made two approximations:
the choice of the classical path for the center of mass trajectory
and the modified area law. 

By the restriction to a straight and steady movement of the center of 
mass $R(\gamma)$ fluctuations of the  bound state 
as a unit are excluded. This approximation is well-founded for heavy bound states.
As we have seen, the typical scale of the problem is the string tension.
The masses of the bound states are  \,$\calM \,\gsim\,3 \sqrt{\sigma}$\,
(cf.\ fig.~\ref{AbbTermschema}) and hence larger than this scale.
It might, nevertheless, be interesting to check this approximation or
to calculate the next order corrections.
Note that the relative movement of both fundamental scalars is treated
fully relativistically in spite of the restriction to the center of mass 
path. 

In this context the obvious question arises why it is possible to 
reduce the treatment of the relative movement to a Schr\"odinger equation. 
Here it is important to note that it is only an analogous non-relativistic 
problem which is described by this equation. The parameters $\mu$ and $\bar\mu$ then  
have the meaning of constituent masses. They are averaged by the 
Schwinger-proper-time integral later on. The evaluation of this integral
with the saddle point method is exact in the 
limit \,$\Theta\!\ra\!\infty$\, investigated by us. 
Neither the non-relativistic treatment of the analog problem nor
the saddle point method is based on any approximations. 
One may, at most, ask if the limit $\Theta$ to infinity can be extrapolated 
by the lattice investigations we are comparing with. 

The modified area law is included in  
our calculations of the 
correlation masses. The correlation with the usual area law was already 
discussed in section \ref{BerKorrMass}. 
An extension to a potential which depends on all three components of $u$ 
does not cause any problem, is, however, not adequate since the connection 
with the scalar-scalar-potential calculated on the lattice would get lost.

\begin{figure}[t]
\begin{picture}(14.0,7.5)
\put(1.0,6.2){$\nfr{V(r)}{g_3^2}$}
\put(7.2,6.1){$\psi_{\rm 1p}(r)$}
\put(7.2,5.05){$\psi_{\rm 1s}(r)$}
\put(12.5,5.5){$\epsilon_{\rm 1p}$}
\put(12.5,4.54){$\epsilon_{\rm 1s}$}
\put(1.2,-0.2){\epsfxsize12cm \epsffile{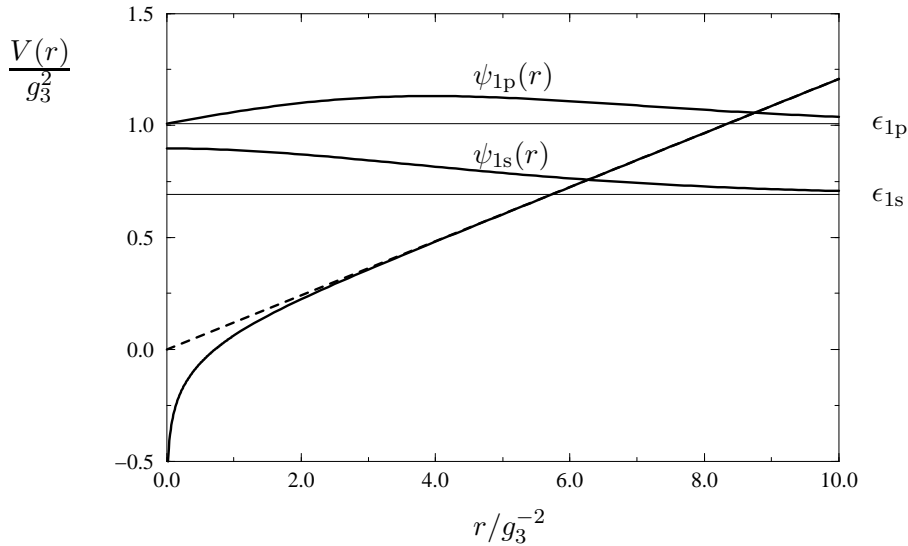}}
\put(7.2,0.1){$r/g_3^{-2}$}
\end{picture}
\caption{The scalar-scalar-potential at \tGl{\ladr}{0.0957} and 
\tGl{\mdrsq}{0.0003} in comparison with the wave functions of the lightest 
bound states.}
\label{AbbPotwithStates}
\end{figure}

Finally we neglected the effect of the screening. 
The Wegner-Wilson-loops calculated on the lattice show no screening of
the potential up to distances of about $8g_3^{-2}$.
In figure \ref{AbbPotwithStates} we plotted $V(r)$ in comparison with
the bound state wave functions.
One sees that the wave functions corresponding to the lowest 
quantum numbers have extensions between $8g_3^{-2}$ and $10g_3^{-2}$.
To exclude screening effects at all the linear part of the potential 
has to be confirmed at distances which are slightly larger than 
those of todays data. 
It is, on the other hand, not expected that the lowest excitations
are influenced by screening effects.

The very good agreement of the correlation masses calculated by us 
with the lattice data justifies our approximations a posteriori. 
As long as the errors of the lattice data do not get 
smaller than today 
we see no need to go along without these assumptions. 

A more interesting question is the calculation of the scalar-scalar-potential.
The Bessel-function contribution \eq{FitK} seems to originate in the exchange 
of a colored massive particle or quasi-particle. The mass is of
order $g_3^2$; the nature of the exchanged particle is not yet revealed.
As explained in appendix \ref{msv}
the string tension and the intercept can in principle be calculated 
with the model of the stochastic vacuum.
The gluon correlator is an important input function for this model,
which needs confirmation from the lattice. 

Another interesting point, which is not fully clarified yet, 
is the treatment of negative values of $m_3^2$. It is
not attractive to have a fundamental scalar with negative squared mass.
One could try to use the perturbative mass calculated from the second
derivative of the effective potential at \tGl{\ph}{0} instead.
The latter one diverges, however, due to the $\ph^2\ln(\ph)$-term 
of the potential. This infrared-divergence shows the breakdown of 
the perturbation theory in the symmetric phase. 
A much more speculative possibility is the generation of a scalar mass
by a gluon-condensate, similar to the generation of a gauge boson mass
by a scalar condensate in the broken phase. 
A gluon condensate due to the instability of the naive vacuum for small
Higgs vev's should also be very important for the generation of the 
non-perturbative part of the electroweak effective potential.

An important outcome of this investigation is that the hot electroweak phase is
strongly interacting  like a pure 3-dimensional Yang-Mills-theory. 
The spectrum of gauge invariant states is, however, totally different from
the latter one. The existence of low-lying 1s- and 1p-states is only due
to the fundamental scalars; the gauge invariant spin~0- and spin~1-operators 
of \eqn{interpolfield}  do not exist in pure Yang-Mills-Theories.
The gauge bosons in the pure gauge theory are 
only defined in a definite gauge while the gauge-invariant spin~1-object
is a W-ball.   
Indeed there is the interesting observation from 
lattice calculations \cite{PhilipsenEA,Philipsen EA2}
that glueballs and Higgs states are practically decoupled.
The string tension is universal in a given confining theory and thus the former
remark fits very well together with the fact mentioned above that the 
string tension responsible for $\bm{\Phi}$-binding is nearly the one  
of pure Yang -Mills theory in lattice calculations.

\section*{Acknowledgement}
We would like to thank M.~Ilgenfritz, M.~Laine, O.~Philipsen and M.~Rueter for 
useful discussions.

\appendix

\section{Perturbative contributions to the potential}\label{pertcont}

Using the cumulant expansion \cite{vKampen} 
the expectation value of the rectangular Wegner-Wilson-loop 
can be expressed as
\be
W(r,T) \gl
{\rm Tr}\; \exp\!\( \,-\; \fr{g_3^2}{2} \; \int\!dx_i \int\!dy_j \;
\langle\langle\, \bm{A}_i(x)\,\bm{A}_j(y) \,\rangle\rangle
\;\;\; +\;{\rm higher~cumulants}\; \)
\quad.
\ee
To lowest order in this expansion one has
\be
\ln \;W(r,T) \gl
-\; \fr{g_3^2}{4} \; \int\!dx_i \int\!dy_j\;
\langle\langle\, A^a_i(x)\,A^a_j(y) \,\rangle\rangle
\quad.
\ee
The perturbative contribution is now calculated from the
exchange of a perturbative gauge boson between the two
long sides of the loop.
With $\vec x$ and
$\vec y$ as two-dimensional vectors and $x_1$ and $y_1$
the corresponding components running along the
long sides one finds
\bea
\lim_{T\to\infty} \,\fr{1}{T}\; 
\ln \;W(r,T) &=&
-\; \fr{g_3^2}{4} \; 
\lim_{T\to\infty} \,\fr{1}{T}\;
\int\limits_{-T/2}^{T/2}\!\!dx_1 \int\limits_{T/2}^{-T/2}\!\!dy_1\;
\langle\langle\, A^a_1(x_1,\vec x)\,A^a_1(y_1,\vec y) \,\rangle\rangle \\
&=&
\fr{g_3^2}{4} \;  \int_{-\infty}^{\infty}\!\!dy_1 \;
\langle\langle\, A^a_1(0,\vec x)\,A^a_1(y_1,\vec y) \,\rangle\rangle
\quad,
\eea
where we have used translational invariance in the last line.

The two-point-cumulant is now replaced by the perturbative 
propagator in Feynman-gauge via
\be
\langle\langle\, A(x)^a_i\,A(y)^b_j \,\rangle\rangle
\;\;\rightarrow\;\; 
\delta^{ab}\;\delta_{ij}\;
\int\!\fr{d^3k}{(2\pi)^3} \; \fr{1}{k^2+m^2} \; 
e^{ik(y-x)}
\quad.
\ee
We split the momentum coordinates in a first component $k_1$
and a two dimensional vector $\vec k$.
The perturbative part of the potential is
\bea
V_{\rm pert}(\vec{y}-\vec{x})
&=&
\fr{g_3^2}{4} \;  \int_{-\infty}^{\infty}\!\!dy_1
\int\!\fr{d^3k}{(2\pi)^3} \; \fr{1}{k_1^2+\vec k^2 +m^2} \; 
e^{i( k_1 y_1 + \vec{k}(\vec{y}-\vec{x}))}\\
&=&
\fr{g_3^2}{4} \; 
\int\!\fr{d^2k}{(2\pi)^2} \; \fr{1}{\vec k^2+m^2} \; 
e^{i\vec{k}(\vec{y}-\vec{x})}
\quad.
\eea
It is thus the two-dimensional Fourier-transformation of
the perturbative propagator in momentum space with \tGl{k_1}{0}.
The derivation holds for \tGl{m}{0} as well.

\section{The model of the stochastic vacuum in three dimensions}\label{msv}

We introduced a model of bound states in the hot 
electroweak phase and showed how to calculate the correlation masses 
within this model. 
The knowledge of the scalar-scalar-potential $V(r)$ is an important 
requirement of this calculation.
The origin of the asymptotically 
linear potential itself has not been explained  
so far. This can be attempted by
adopting the model of the stochastic vacuum, which was originally proposed 
by Dosch and Simonov \cite{Dosch1,DoschSi} for QCD, 
to the three dimensional SU(2)-Higgs-model.

The basic ingredient of the model of the stochastic vacuum is the
assumption that the complicated contributions of non-perturbative field
configurations can be approximated by a simple stochastic process. In
that way already the assumption that this process has a convergent linked
cluster expansion leads very naturally to linear confinement. Making
the more restrictive approximation of a Gaussian stochastic process
leads to a very predictive model since now in principle the full
non-perturbative contribution is approximated by a single correlator. 
This correlator can be used to determine observables as the static
quark-antiquark potential in QCD, soft high energy cross sections and others.
It can also be compared with lattice calculations and all results
turned out to be very satisfactory. 

In the following we shortly exhibit basic features of the model adopted
to three dimensions and SU($N_c$) as gauge group with \tGl{N_c}{2}, 
the case considered here.
In three dimensions we can introduce besides the field tensor 
$F^a_{ij}$ the vector 
\be
\tilde F^a_k \gl \epsilon_{ijk}\,F_{ij}^a
\ee
which in an Abelian theory satisfies the Bianchi identity
\be\label{Bianchi}
\del_k \tilde F^a_k = 0
\quad.
\ee
In order to form gauge 
invariant correlators of the field strength tensors  we have to
transport the color content of all 
fields at point $x$  to a fixed reference point $y$. 
This is done by the gauge field transporter $\bm{T}(x,y)$ \eq{linkop} 
along a straight line from $x$ to $y$.
\be\label{transportedF}
F^a_{ij}(x,y)\;\fr{\tau^a}{2} \gl 
\bm{T}(x,y)^{-1}\, F^a_{ij}(x)\;\fr{\tau^a}{2}\, \bm{T}(x,y)
\quad.
\ee

The correlator 
\,$\langle\, g_3^2 F^a_{ij}(x,y) F^b_{kl}(x',y)\,\rangle_{\!\!\bmind{A}}$\,
depends in general on the reference point $y$. We make the
assumption  that it can be approximated by an expression
depending only on the difference \tGl{z}{x-x'}.
\bea\label{FxFy}
\langle\, F^a_{ij}(x,y) F^b_{kl}(x',y) \,\rangle_{\!\!\bmind{A}}
&=&
\fr{\delta^{ab}}{6(N_c^2\!-\!1)} \;\langle FF \rangle\;
\Bigg[\; \kappa \, (\delta_{ik}\delta_{jl} - \delta_{il}\delta_{jk}) \, 
    D(z^2) \\
&& \!\!\mquad +\;
 (1\!-\!\kappa)\left(\frac{1}{2}\frac{\partial}{\partial z_i} 
 (z_k \delta_{jl} - z_l \delta_{jk} ) + 
      \frac{1}{2}\frac{\partial}{\partial z_j} (z_l \delta_{ik} - 
   z_k \delta_{il}) \right) D_1(z^2) \Bigg]
\quad.\nonumber
\end{eqnarray}
Here $\langle FF \rangle$ is the gluon condensate 
\be
\langle FF \rangle  \gl  
\sum_{ij}\sum _a \; \langle\, F^a_{ij}(0) F^a_{ij}(0) \,\rangle_{\!\!\bmind{A}}
\ee
and $D$  and $D_1$ are correlation functions which are supposed to fall
off with a characteristic correlation length $a$.  

In an Abelian gauge (without monopoles) the Bianchi identity \eq{Bianchi} 
forces the first contribution to vanish, i.e.\ \tGl{\kappa}{0}, 
but in a non-Abelian theory we have no longer the Bianchi identity 
and hence there is no
reason that the correlator $D(z^2)$ vanishes. 
Lattice results \cite{DiGP92} showed $\kappa\approx 0.74$ in QCD.
No lattice results are available for the 3-dimensional correlator.

From the correlator one can obtain the area law 
by first transforming the line integral over the potential $A^a_i$ into
a surface integral over the field strength $F^a_{ij}$  by means of
the non-Abelian Stokes theorem and then applying the cluster expansion.
For subtleties due to the path ordering we refer to reference \cite{DoRue}.

For a rectangular Wegner-Wilson-loop of extension $T$  
in $z_0$-direction and $r$ in $z_1$-direction we obtain:
\begin{eqnarray}
\ln W(r,T) &=&  \ln N_c \;-\; \frac{g^2\langle FF \rangle}{6 N_c}
\left[\; \int_0^T\!\!d z_0 \int_0^r\!\!dz_1\; 
\kappa\, (r T - r z_0 - T z_1 + z_0 z_1)\, D(z^2)
\right. \nonumber  \\ 
&& \left. +\;
(1\!-\!\kappa)\, \(\fr12 T z_1 + \fr12 r z_0 - z_1 z_0 \) \, D_1(z^2) \;\right]
\end{eqnarray}
For small values of $r$ and $T$ the non-constant term in \,$\ln W$
behaves as  \,$- \tfr1{24 N_c} g^2\langle FF \rangle\, r^2T^2 $, 
independent of the value of $\kappa$, whereas for large values of $r$ and $T$ 
only the correlator $D(z^2)$ gives an area law. 

The static potential is obtained from $W(r,T)$ by the limit of \eqn{DefV}.
It increases quadratically in $r$ for \,$r \!\ll\! a$\, and 
linearly for \,$r \!\gg\! a$\, (cf.\ fig.~\ref{AbbModlinPot}b).
Asymptotically, i.e.\ for \,$r \!\gg\! a$\,, the potential $V(r)$ can 
be written in the form of \eqn{Vlowerd}.
The string tension is determined by the gluon condensate and the correlation
length via \eqn{stringtension}.
The value of the intercept $V_0$ depends on the value of $\kappa$ and the form
of the correlators $D$ and  $D_1$. For $\kappa=1$ one 
obtains \eqn{interceptmsv}.

Therefore the linear rising of the potential at large distances as
well as a negative intercept can be explained within the model of the 
stochastic vacuum. 
Note that the model of the stochastic vacuum explains only the 
non-perturbative part of the potential. The perturbative part has to
be added.
In the three-dimensional case  
it is, however, not easy to distinguish from the non-perturbative 
part because the squared gauge coupling has dimension of 
a mass.

\end{document}